\documentclass[11pt]{article}

\usepackage[sort,longnamesfirst]{natbib}

\usepackage{amsbsy,amsmath,amsthm,amssymb,graphicx,verbatim,color}
\usepackage{subfigure}
\usepackage{multirow}

\usepackage{geometry}
{\begin{list}{}%
         {\setlength{\leftmargin}{#1}}%
         \item[]%
}
{\end{list}}

\usepackage{multirow}
\usepackage{amsbsy,amsmath,amsthm,amssymb,verbatim,color}
\usepackage{graphics}
\usepackage{graphicx,subfigure}
\usepackage{float,graphicx}
\usepackage{geometry}
\usepackage{setspace}
\usepackage[all]{xy}

\newtheorem{theorem}{Theorem}
\newtheorem{proposition}{Proposition}

\theoremstyle{remark}

\newcommand{\lp}{\left(}
\newcommand{\rp}{\right)}
\newcommand{\lb}{\left\lbrace}
\newcommand{\rb}{\right\rbrace}

\newcommand{\sig}{\sigma}
\newcommand{\lt}{\lambda_\theta}
\newcommand{\lame}{\lambda_e}

\newcommand{\lam}{\lambda}
\newcommand{\ooy}{\overline{\overline{y}}}
\newcommand{\vareps}{\varepsilon}

\begin{document}

%computing journal:  journal of stat \& computing, CSDA, canadian journal

\title{A Modified Gibbs Sampler on General State Spaces}

\author{ Alicia A. Johnson\footnote{ {\tt ajohns24@macalester.edu}}  \\
            Department of Mathematics, Statistics, and Computer Science \\
            Macalester College \and
James M. Flegal\footnote{{\tt jflegal@ucr.edu, Research supported by the National Science Foundation}} \\
Department of Statistics \\ University of California, Riverside
       }
       \date{} 
\maketitle

\begin{abstract}

We present a modified Gibbs sampler for general state spaces.  We establish that this modification can lead to substantial gains in statistical efficiency while maintaining the overall quality of convergence.  We illustrate our results in two examples including a toy Normal-Normal model and a Bayesian version of the random effects model.   
\end{abstract}

\newpage

\section{Introduction}

%Consider random variable $X=(X_1,\ldots,X_m)$ having probability distribution $\varpi$ with support on $ \mathcal{X} = \mathcal{X}_1 \times \cdots \times \mathcal{X}_m$ where $X_i \in \mathbb{R}^{d_i}$ and $\mathcal{X}_i \subseteq \mathbb{R}^{d_i}$ for $d_i \ge 1$. 

Consider a random variable $X=(X_1,\ldots,X_m)$ where $X_i \in \mathbb{R}^{d_i}$ for $i=1,\ldots, m$ and $d_i \ge 1$. Let $X$ have probability distribution $\varpi$ with support $\mathcal{X} = \mathcal{X}_1 \times \cdots \times \mathcal{X}_m$ and associated conditional distributions $\varpi_{X_i|X_{-i}}$ where $X_{-i}=X \backslash X_i$.
When $\varpi$ is intractable, inference regarding $X$ may require Markov chain Monte Carlo (MCMC) methods.  
To this end, 
consider using the Gibbs sampling algorithm (GS) under a random scan to construct a Markov chain denoted
\[
\Phi = \lb X^{(0)}, X^{(1)}, \ldots \rb = \lb \left(X_1^{(0)},\ldots,X_m^{(0)}\right), \left(X_1^{(1)},\ldots,X_m^{(1)}\right), \ldots \rb \; .
\]
Under a fixed set of probabilities  $p=(p_1,\ldots,p_m)$ where $0 < p_i < 1$ and $\sum_{i=1}^m p_i=1$, $\Phi$ moves from 
$X^{(i)}=x$ to $X^{(i+1)}$ by updating a single randomly selected $X_i$ while fixing all others.
Specifically, iteration $j+1$ of the GS first draws $(Z_1,\ldots,Z_m) \sim \text{Multinomial}(1,p)$.  Then for $\{ i : Z_i=1 \}$, draw $x_i' \sim \varpi_{X_i|X_{-i}}(\cdot | x_{-i})$ and set 
\[
\left(X_1^{(j+1)},\ldots,X_m^{(j+1)}\right)=(x_1, \ldots, x_{i-1}, x_i', x_{i+1}, \ldots,x_m) \; .
\]

%\vspace{.15in}
%\hrule
%\vspace{.1in}
%
%Iteration $i+1$ of the GS:
%
%\begin{enumerate}
%\item Draw $Z=(Z_1,\ldots,Z_m) \sim \text{Multinomial}(1,p)$.
%\item If $Z_1=1$: Draw $x_1' \sim \varpi_{X_1|X_{-1}}(\cdot | x_{-1})$ and set $\left(X_1^{(i+1)},\ldots,X_m^{(i+1)}\right)=(x_1',x_2,x_3,\ldots,x_m)$.
%
%If $Z_2=1$: Draw $x_2' \sim \varpi_{X_2|X_{-2}}(\cdot | x_{-2})$ and set $\left(X_1^{(i+1)},\ldots,X_m^{(i+1)}\right)=(x_1,x_2',x_3,\ldots,x_m)$.
%
%
%$\vdots$
%
%
%If $Z_m=1$: Draw $x_m' \sim \varpi_{X_m|X_{-m}}(\cdot | x_{-m})$ and set $\left(X_1^{(i+1)},\ldots,X_m^{(i+1)}\right)=(x_1,\ldots,x_{m-1},x_m')$.
%\end{enumerate}
%
%\vspace{.15in}
%\hrule
%\vspace{.1in}

After $n$ iterations, we can estimate the expected value $\beta := E_\varpi  f = \int f(x) \varpi(dx)$ of some function of interest, $f: \mathcal{X} \to \mathbb{R}$,
by Monte Carlo average
$\hat{\beta}_n := \frac{1}{n}\sum_{i=0}^{n-1} f\left( X^{(i)}\right)$.
The level of confidence we can place in $\hat{\beta}_n$ is intimately tied to the rate at which $\Phi$ converges to $\varpi$.
%
% any Markov chain algorithm is bolstered by [1] quick convergence to $\varpi$; and [2] efficiency.
%Consider [1] and define $n$-step transition kernel
To this end, assume $\Phi$ is Harris ergodic \citep{meyn:twee:1993} and define $n$-step transition kernel
$P^n(x,A) = \text{Pr}\left(X^{(i+n)} \in A \; \vline \; X^{(i)} = x \right)$ for $x \in \mathcal{X}$, $n,i \in \mathbb{N}$, and $A \in \mathcal{B}$
where $\mathcal{B}$ is the Borel $\sigma$-algebra associated with $\mathcal{X}$.
Then we say $\Phi$ is {\it geometrically ergodic} if it converges to $\varpi$ in total variation distance at a geometric rate.  That is, there exist function $M : \mathcal{X} \to \mathbb{R}$ and constant $t \in (0,1)$ such that
\[
\parallel P^n(x, \cdot) - \varpi(\cdot) \parallel_{\text{TVD}} \; :=  \sup_{A \in \mathcal{B}} | P^n(x,A) - \varpi(A)| \le t^n M(x) \;\;\; \text{ for all } x \in \mathcal{X}\; .
\]
In addition to guaranteeing effective simulation results in finite time, geometric ergodicity is a key sufficient condition for the existence of a Markov chain Central Limit Theorem for $\hat{\beta}_n$ \citep{jone:2004}.
%the following Markov chain Central Limit Theorem (CLT) \citep{jone:2004}.   
%If $\Phi$ is geometrically ergodic and $E_\varpi[f(X)]^{2+\delta} < \infty$ for some $\delta>0$, then
%\[
%\sqrt{n}(\hat{\beta}_n - \beta) \stackrel{D}{\to} N(0, v(f,P))
%\]
%where asymptotic variance
%$v(f,P) = Var_\varpi[f(X)] + 2\sum_{i=1}^\infty Cov\left[ f\left(X^{(0)}\right),  f\left(X^{(i)}\right)\right]$
%depends both on function $f$ and the structure of transition kernel $P$.   

Inspired by the work of \cite{liu:1996b} on discrete state spaces, 
we show that a simple modification to the GS can lead to significant improvements in Markov chain efficiency and quality of estimates $\hat{\beta}_n$.
Specifically, we introduce a conditional Metropolis-Hastings algorithm (CMH) that increases efficiency by encouraging movement of $X_j^{(i+1)}$ outside the local neighborhood of $X_j^{(i)}$, denoted $B_j \subset \mathcal{X}_j$.
We show that this modification maintains the overall quality of convergence;
geometric ergodicity of the CMH guarantees the same for the GS and, under conditions on $B_j$, the reverse is also true.

Further, we explore the impact of $B_j$ on the CMH and compare the empirical performance of the GS and CMH in two different model settings: (1) the Normal-Normal model; and (2) a Bayesian version of the random effects model.   The latter is practically relevant in that inference for this model {\it requires} MCMC methods. 
In both settings, the CMH with reasonably sized $B_j$ is significantly more efficient than the GS in both its movement around state space $\mathcal{X}$ and in its estimation of expected value $\beta$.  However, there are limits to the CMH efficiency.  Mainly, when the $B_j$ are too large, the CMH is pushed out to the `edges' of the state space and cannot compete with the GS.

The following section introduces the CMH and compares convergence among the GS and CMH.  Later, Section \ref{sec:example} explores our results with applications in two model settings. All proofs are deferred to the appendix.

\section{Conditional Metropolis-Hastings Modification}\label{sec:GS}

\subsection{The CMH Algorithm}
Suppose $\varpi$ admits density $\pi(x_1,\ldots,x_m)$ with respect to measure $\mu = \mu_1 \times \cdots \times \mu_m$.  Further, let $\pi(x_i|x_{-i})$ denote the associated full conditional densities from which we assume direct simulation is possible. 
In each iteration, the GS for $\varpi$ updates a single randomly selected $X_i$ conditional upon the current values of all other $X_j$.  
Thus the GS transition kernel can be expressed as 
\[
P_{GS}(x,A)  = \sum_{i=1}^m p_i P_{GS_i}(x_{-i}, A) 
\]
where the $P_{GS_i}$, Markov kernels corresponding to the $X_i$ updates, are defined by  
\[
P_{GS_i}(x_{-i},A) = \int_{\{x_i': (x_{[i-1]},x_i',x^{[i+1]}) \in A\}} \pi(x_i' | x_{-i}) \mu_i(dx_i') 
\]
for $x_{[i]} = (x_1,\ldots, x_i)$ and $x^{[i]} = (x_i,\ldots,x_m)$.
%& = \sum_{i=1}^m p_i \int_{\{x_i': (x_{[i-1]},x_i',x^{[i+1]}) \in A\}} k_{GS_i}(x,x') \mu_i(dx_i')  \\
%& =\int_{CMH} k_{CMH}(x,x') \mu(dx')  \\
%$P_{GS}(x,A) = \int_{CMH} k_{GS} (x,x') \mu(dx')$
%for Markov transition density (Mtd) 
%\[
%k_{GS}(x,x') = \sum_{i=1}^m p_i \pi(x_i'|x_{-i})\delta(x_{-i}-x_{-i}')
%\]
%where $\delta$ is Dirac's delta.  
Ideally, the GS will tour all reaches of $\mathcal{X}$ without getting stuck for too long in any one ``corner."  Indeed, we can facilitate such movement with a simple modification to the GS algorithm.  Letting $x$ denote the current state of the GS, suppose component $x_i$ is selected for update and let $B_i(x_i|x_{-i}) \subset \mathcal{X}_i$ be a local neighborhood of $x_i$ that could depend on $x_{-i}$.  For example, we might define $B_i(x_i|x_{-i}) = x_i \pm \varepsilon$ for $\varepsilon > 0$ when $\mathcal{X}_i = \mathbb{R}$ or define $B_i(x_i|x_{-i})$ to be a circle centered at $x_i$ with radius $\varepsilon$  when $\mathcal{X}_i = \mathbb{R}^2$. 
Then instead of drawing an $x_i$ update from  $\pi(\cdot|x_{-i})$, we can restrict movement to states outside $B_i(x_i|x_{-i})$, i.e..~$B_i^c(x_i|x_{-i}) = \mathcal{X}_i \backslash B_i(x_i|x_{-i})$, through a Metropolis-Hastings step as follows.  First draw $x_i'$ from the proposal density
\[
q_i(x_i'|x) = \frac{\pi(x_i'|x_{-i})}{\int_{ B_i^c(x_i|x_{-i})} \pi(z_i|x_{-i}) \mu_i(dz_i)}I(x_i' \in B_i^c(x_i|x_{-i}))
\]
using a simple accept-reject strategy,  then replace $x_i$ with $x_i'$ with acceptance probability
\[
\alpha_i(x_i'|x) = \min \left\lbrace 1, \; \frac{\int_{ B_i^c(x_i|x_{-i})}\pi(z_i|x_{-i}) \mu_i(dz_i)}{\int_{ B_i^c(x_i'|x_{-i})}\pi(z_i|x_{-i}) \mu_i(dz_i)}\right\rbrace \; .
\]
Thus, the CMH modification of the GS has transition kernel
\[
P_{CMH}(x,A) = \sum_{i=1}^m p_i P_{CMH_i}(x, A)\\
\]
for
\[
P_{CMH_i}(x,A) 
 = \int\limits_{\{x_i': (x_{[i-1]},x_i',x^{[i+1]}) \in A\}} \hspace{-.4in} q_i(x_i'|x) \alpha_i(x_i'|x) \mu_i(dx_i')   + \left[1 - \int q_i(x_i'|x) \alpha_i(x_i'|x) \mu_i(dx_i')  \right] I(x\in A) \; . \\
\]
Note the dependence of the CMH on neighborhoods $B_i$.  If $B_i(x_i|x_{-i}) = \emptyset$ for all $i$,  the CMH and GS are equivalent.  At the other extreme, when $B_i(x_i|x_{-i}) = \mathcal{X}_i$, the CMH Markov chain has nowhere to move.  Thus we restrict our attention to the CMH with
  \begin{equation}\label{eq:nbhd}
\begin{split}
%& \int_{ B_i(x_i|x_{-i})} \pi(z_i|x_{-i}) \mu_i(dz_i) >0 \text{ for at least one $i\in \{1,\ldots,m\}$; } \hspace{.1in} \text{ and }\\
& \sup_{x \in \mathcal{X}, \; i\in \{1,\ldots,m\}}\int_{ B_i(x_i|x_{-i})} \pi(z_i|x_{-i}) \mu_i(dz_i)  < 1 \; .\\
\end{split}
\end{equation} 
Of course, there are countless ways to select $B_i(x_i|x_{-i})$ that satisfy \eqref{eq:nbhd}.  To an extent, this choice is context dependent and presents a goldilocks challenge.  When the $B_i(x_i|x_{-i})$ are too small relative to $\mathcal{X}_i$, the boost in efficiency is not great enough to warrant the CMH modification.  When the $B_i(x_i|x_{-i})$ are too large, the CMH gets trapped exploring the edges of $\mathcal{X}$ with low probability $\alpha_i(x_i'|x)$ of getting `unstuck.'
Further, the typical computation time required for an accept-reject draw of $x_i'$ outside $B_i(x_i|x_{-i})$ increases.  
%Indeed, we present an example in Section \ref{sec:example} which suggests there are limits to the improvement in efficiency as $\mu_i(B_i(x_i|x_{-i}))$ increases.
Thus the selection of $B_i(x_i|x_{-i})$ is a delicate issue.
Though not the focus of this preliminary study, we discuss this choice in more detail in Section \ref{sec:example}.

\subsection{Convergence}

While facilitating movement around the state space, the CMH modification to the GS can be constructed in a way that also preserves the overall quality of convergence.
%Under limits on the size of neighborhoods $B_i(x_i|x_{-i})$, convergence properties of the GS are passed along to its CMH modification.
%the CMH is guaranteed to enjoy convergence properties similar to that of the GS. 
To this end, we require a few definitions.  
Suppose Markov chain $\Phi$ with transition kernel $P$ is Harris ergodic and Feller, i.e.~for any open set $O \in \mathcal{B}$ and $x', x \in \mathcal{X}$,
$\liminf_{x' \to x} P(x', O) \ge P(x, O)$.  Then $\Phi$ is geometrically ergodic if and only if there exists some {\it drift function} $V: \mathcal{X} \to [1,\infty)$ such that $V$ is {\it unbounded off compact sets} (i.e.~$\{x: V(x) \le d\}$ is compact for all $d > 0$) and satisfies  {\it drift condition} 
\begin{equation}\label{eq:drift}
PV(x) :=  \int V(x') P(x, dx') \le \gamma V(x) + b \;\;
\end{equation}
for  $x \in \mathcal{X}$, constant $b < \infty$, and {\it drift rate} $0 < \gamma <1$ where smaller $\gamma$ are loosely indicative of quicker convergence \citep[see, for example,][]{jone:hobe:2001}. 
In light of these properties, we can derive conditions under which the CMH inherits geometric ergodicity from the GS. 
\begin{theorem}\label{thm:gscmh}
Consider the GS and CMH for $\varpi$ and assume both are Harris ergodic and Feller.  Further, suppose the CMH has neighborhoods $B_i(x_i|x_{-i})$ satisfying \eqref{eq:nbhd} 
with 
\[
q_{\min} \le \int_{B_i(x_i|x_{-i})} \pi(z_i|x_{-i}) \mu_i(dz_i) \le q_{\max} \; \text{ for all } x \in \mathcal{X}, \; i \in \{1,\ldots,m\}
\]
where $0 \le q_{\min} \le q_{\max} < 1$.
Then if the GS is geometrically ergodic with drift condition \eqref{eq:drift}, 
the CMH is geometrically ergodic so long as 
%at least one of the following holds:
%\begin{enumerate}
%\item $(1-q_{\max})(1-q_{\min})^{-1} > \gamma$ and 
%there exists $c \in \mathbb{R}^+$ such that
%\[
%\sup_{x,x' \in \mathcal{X}, i\in \{1,\ldots,m\}} \; \lb \; \vline \; V(x) - V\left(x_{[i]},x_i',x^{[i+1]}\right) \; \vline I(x_i' \in B_i(x_i|x_{-i}))\rb \le c \;  ; \; \text{ OR } 
%\]
%\item 
\begin{equation}\label{eq:q}
q_{\max} < \frac{1}{2}
\hspace{.2in} \text{ and } \hspace{.2in} 
\frac{1-2q_{\max} + q_{\min}q_{\max}}{1-q_{\min}} > \gamma \; .
\end{equation}
%\end{enumerate}
\end{theorem}

It is important to note that drift condition \eqref{eq:drift} is not unique.  Thus the restrictions \eqref{eq:q} on $q_{\max}$, $q_{\min}$ can be overly conservative.  However,  the requirement that $q_{\max}$ be less than 1/2 is consistent with our intuition.  Consider a simple example.

\vspace{.1in}
{\it Example.}  Let $(X_1,X_2)$ be uniform on the unit square with $\pi(x_1,x_2) = 1$ and $\pi(x_i|x_{-i}) = 1$  for $(x_1,x_2) \in [0,1]^2$.
Consider the CMH for $\pi$ with $B_i(x_i|x_{-i}) = x_i  \pm \varepsilon/2$ for $\varepsilon \in (0,1)$.  In this case, $q_{\max} = \varepsilon$.  Further, when $q_{\max} = \varepsilon  > 1/2$, the CMH is restricted to movement outside the center square $[1-q_{\max}, q_{\max}]^2$.  Thus, with the starting value as the only exception, the CMH will never visit the middle $(2q_{\max} - 1)^2$ proportion of target uniform distribution.

\vspace{.1in}
The restrictions \eqref{eq:q} also provide interesting insight into the challenge of selecting $B_i(x_i|x_{-i})$.  Mainly, the larger the GS drift rate $\gamma$, the smaller $q_{\max}$ must be.  Thus when the geometric convergence rate of GS is slow, only the CMH with small neighborhoods $B_i(x_i|x_{-i})$ (i.e.~those making small modifications to the GS) are guaranteed to inherit geometric ergodicity.
%These limitations lessen when the GS converges quickly.  
In contrast, geometric ergodicity of {\it any} CMH guarantees the same for the GS.

\begin{theorem}\label{thm:subgeo}
If the CMH with neighborhoods $B_i(x_i|x_{-i})$ satisfying \eqref{eq:nbhd} is geometrically ergodic, the GS is also geometrically ergodic. 
\end{theorem}

\section{Examples}\label{sec:example}
%show that, by facilitating movement around the state space, the CMH modification to the GS can improve efficiency while maintaining the overall quality of convergence.  Also goldilocks.

We present applications of the GS and CMH in two different model settings: (1) the Normal-Normal model; and (2) a Bayesian version of the random effects model.    In both cases, we compare the empirical performance of the finite sample GS and CMH with respect to the following measures of efficiency.

The {\it expected square jump distance} (ESJD) provides a measure of how efficiently the CMH and GS traverse $\mathcal{X}$.  Specifically, ESJD is the expected value of the average squared distance that a chain travels in a single iteration.  To estimate the CMH ESJD  (similarly for GS), we run $N$ independent CMH chains of length $n$ and calculate the sample mean squared jump distance for each:  
\[
\text{MSJD}_{CMH}^{(i)} = \frac{1}{n-1}\sum_{j=1}^{n-1} || X^{(j+1)} - X^{(j)} ||^2_2 \;\; \text{ for } i \in \{1,\ldots,N\}
\]
where $|| \cdot ||_2$ denotes the Euclidean norm.  Thus an estimate of the CMH ESJD is calculated by
\[
\widehat{\text{ESJD}}_{CMH} = \frac{1}{N}\sum_{i=1}^N \text{MSJD}_{CMH}^{(i)} \; .
\]
In turn, the relative efficiency of the CMH and GS in exploring $\mathcal{X}$ can be estimated by the ratio  $\widehat{\text{ESJDR}} = \widehat{\text{ESJD}}_{CMH}/\widehat{\text{ESJD}}_{GS}$.

We also want to compare the efficiency of the CMH and GS relative to the estimation of $\beta = E_\varpi \left[ f(X) \right]$ by 
Monte Carlo averages $\hat{\beta}_{CMH,n}$ and $\hat{\beta}_{GS,n}$, respectively.  
%When the CMH and GS are geometrically ergodic and $E_\varpi[f(X)]^{2+\delta}<\infty$ for some $\delta>0$, the following CLTs hold as $n \to \infty$:
%%\begin{equation}\label{eq:clt}
%\[
%\sqrt{n}(\hat{\beta}_{CMH,n} - \beta) \stackrel{D}{\to} N(0, v(f,P_{CMH})) 
%\hspace{.2in} \text{ and } \hspace{.2in}
%\sqrt{n}(\hat{\beta}_{GS,n} - \beta) \stackrel{D}{\to} N(0, v(f,P_{GS})) 
%\; .
%\]
%Thus the {\it asymptotic relative efficiency} of the CMH and GS,
%\[
%\text{ARE} = \frac{v(f,P_{CMH})}{v(f,P_{GS})}\;,
%\]
%is a ratio of the Markov chain sample sizes needed for the CMH and GS to achieve the same level of precision in estimating $\beta$.  
%Under geometric ergodicity, a consistent estimator of ARE is provided by $\widehat{\text{ARE}} = \hat{v}(f,P_{CMH})/\hat{v}(f,P_{GS})$ where the $\hat{v}$ are batch mean estimates of $v$ \citep{fleg:jone:2010} calculated from single, length $n$ runs of the CMH and GS.
The combined bias and variance of these estimators is captured by the mean squared error (MSE):
\[
\text{MSE}_{CMH} = \text{E}(\hat{\beta}_{CMH,n} - \beta)^2  
\hspace{.2in} \text{ and } \hspace{.2in} 
\text{MSE}_{GS} = \text{E}(\hat{\beta}_{GS,n} - \beta)^2  
\; .
\]
We estimate $\text{MSE}_{CMH}$ (similarly, $\text{MSE}_{GS}$) using $N$ independent estimates $\lb\hat{\beta}_{CMH,n}^{(1)},\ldots,\hat{\beta}_{CMH,n}^{(N)}\rb$
obtained from $N$ independent CMH chains of length $n$:
\[
\widehat{\text{MSE}}_{CMH} =
\frac{1}{N}\sum_{i=1}^{N}(\hat{\beta}_{CMH,n}^{(i)} - \beta)^2 \; .
\]
In turn, we estimate the MSE ratio of  CMH relative to GS  by $\widehat{\text{MSER}} = \widehat{\text{MSE}}_{CMH} /\widehat{\text{MSE}}_{GS}$.

\subsection{Normal-Normal model}\label{sec:NN}

Let $\pi(x_1,x_2)$ on $\mathbb{R}^2$ denote the density corresponding to bivariate Normal distribution
\[
\left( \begin{array}{c} X_1 \\ X_2 \end{array} \right) \sim N_2 \left(\left( \begin{array}{c} 0 \\ 0 \end{array} \right), \; \left( \begin{array}{cc} 2 & 1 \\ 1 & 1 \end{array} \right) \right)  \; .
\]
Further, let densities $\pi(x_1|x_2)$ and $\pi(x_2|x_1)$ correspond to the associated conditionals
\[
X_1 | X_2 \sim N(X_2, 1) 
\hspace{.1in} \text{ and } \hspace{.1in} X_2 | X_1 \sim N\left(\frac{X_1}{2}, \frac{1}{2}\right) \; .
\]
We compare the GS and CMH for $\pi$, both algorithms starting at $\left(X_1^{(0)},X_2^{(0)}\right) = (0,0)$ and updating $X_1$ and $X_2$ with equal probability in each iteration (i.e.~$p_1=p_2=0.5$).  Further, we use two strategies for selecting CMH neighborhoods $B_i(x_i|x_{-i})$.
First, let $\text{CMH}_c$ denote the CMH with $B_i(x_i|x_{-i})$ encompassing all values within $c>0$ conditional standard deviations of $x_i$:
\[
B_1(x_1|x_2) = x_1 \pm c
\hspace{.2in} \text{ and } \hspace{.2in}
B_2(x_2|x_1) = x_2 \pm c\sqrt{\frac{1}{2}} \; .
\]
Alternatively, let $\text{CMH}_q$ denote the CMH with $B_i(x_i|x_{-i})$ centered at $x_i$ and having fixed density $0 < q < 1$.  That is, $B_i(x_i|x_{-i}) =  x_i\pm d(x_i)$ for $d(x_i)$ that satisfies
\[
\int\limits_{x_i-d(x_i)}^{x_i+d(x_i)}\pi(z_i|x_{-i}) dz_i = q \;.
\]
In this special case, the CMH acceptance probability $\alpha_i(x_i'|x) = 1$ for all $x',x \in \mathcal{X}$.  

The GS, CMH$_c$, and CMH$_q$ are Harris ergodic and Feller.  The GS is also known to be geometrically ergodic, satisfying drift condition 
\[
P_{GS}V(x_1,x_2) = \gamma V(x_1,x_2) + b
\]
for $V(x_1,x_2) = x_1^2 + 2x_2^2$, $\gamma=0.75$, and $b=1$ \citep[see, for example,][]{john:2009}.  Thus Proposition \ref{prop:nn} follows from Theorem \ref{thm:gscmh}. 
\begin{proposition}\label{prop:nn}
$\text{CMH}_c$ is geometrically ergodic for $c < 0.1573$ and $\text{CMH}_q$ is geometrically ergodic for $q <  0.25$.
\end{proposition}

As noted above, since the CMH neighborhood restrictions in Proposition \ref{prop:nn} are derived from a non-unique drift condition, they are likely conservative.  That is, it is probably the case that $\text{CMH}_c$ and $\text{CMH}_q$ are geometrically ergodic for broader ranges of $c$ and $q$.  However, establishing CMH specific drift conditions is more difficult than for the GS.  Thus a more general result eludes us.

\begin{figure}[H]
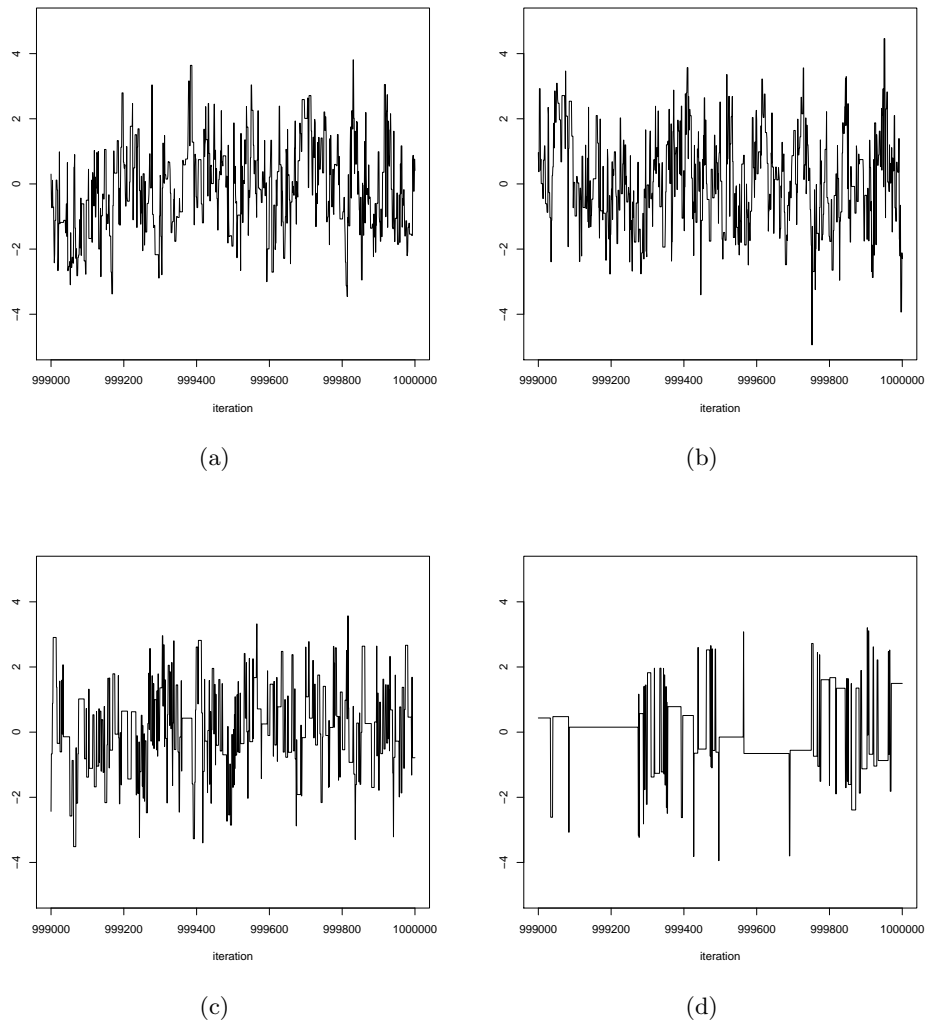
\centering
 \subfigure[]{\includegraphics[height=2.5in,width=2.5in]{trace1.jpg}}
 \subfigure[]{\includegraphics[height=2.5in,width=2.5in]{trace2.jpg}} 
 \subfigure[]{\includegraphics[height=2.5in,width=2.5in]{trace3.jpg}}
 \subfigure[]{\includegraphics[height=2.5in,width=2.5in]{trace4.jpg}}
 \caption{Trace plots of $X_1$ for iterations 9.99e5 through 1e6 of 
 the (a) GS, (b) CMH$_{c=0.1}$, (c) CMH$_{c=1.5}$, and (d) CMH$_{c=3}$ for the Normal-Normal example of Section \ref{sec:NN}.}
 \label{fig1}
\end{figure}

Next, we compare the efficiency of the GS and CMH relative to the estimation of $\beta=E(X_1)=0$.
% by $\hat{\beta}_{GS,n}$ and $\hat{\beta}_{CMH,n}$, respectively.  
%Under geometric ergodicity, CLTs are guaranteed for $\hat{\beta}_{GS,n}$ and $\hat{\beta}_{CMH,n}$ since $E(X_1^4) < \infty$. 
To begin, consider CMH$_c$, the CMH using the fixed width neighborhood strategy.   Table \ref{tab1} presents estimates $\widehat{\text{ESJDR}}$ and $\widehat{\text{MSER}}$
calculated from $N=1000$ independent, length $n=1000$ runs of the GS and CMH$_c$ for each $c \in \{0.1, 0.5, 1, 1.5, 2, 2.5, 3\}$.
In addition, 
estimates of the 
%asymptotic relative efficiency\footnote{Note that Proposition \ref{prop:nn} only guarantees consistency for $\widehat{\text{ARE}}$ when $c < 0.1573$.}, $\widehat{\text{ARE}}$, and 
CMH$_c$ acceptance rates  (i.e.~the percent of M-H proposals accepted) were obtained from independent runs of CMH$_c$, each of length $n=10^6$.  
Overall these results suggest that, for reasonably sized neighborhoods ($c \le 2$), the CMH$_c$ is more efficient than the GS in its movement around $\mathbb{R}^2$ and in its estimation of $\beta$.  Across the board, CMH$_c$ efficiency peaks with $c=1.5$ where the CMH enjoys the largest average per iteration movement and smallest MSE.
%where nearly 1.4 (1/0.72) GS samples are required for each single CMH sample in order to achieve the same precision in estimating $\beta$.
On the other hand, CMH$_c$ efficiency suffers when the neighborhoods are too large ($c>2$).  Mainly, as the neighborhood size increases, acceptance rates decrease and the CMH$_c$ no longer enjoys efficient movement around the state space. In fact, we observe that CMH$_c$ performance is weakest when acceptance rates are closest to the gold standard of 0.234 ($c =2.5, 3$).  Indeed, we expected this to be the case and merely include these results to demonstrate the limits of CMH efficiency.  Let $\left(X_1^{(i)}, X_2^{(i)}\right)$ denote the current state of the CMH chain and suppose $X_1^{(i)}$ is selected for update.  When $c=2.5$ ($c = 3$), the only movement $X^{(i+1)}$ can make is to the extreme upper and lower 0.62\% (0.13\%) of the  N$\left(X_2^{(i)}, 1\right)$ full conditional distribution.  We illustrate this behavior in Figure \ref{fig1}, a comparison of trace plots for the GS, CMH$_{c=0.1}$, CMH$_{c=1.5}$, and CMH$_{c=3}$.

\begin{table}[H] \centering
 \caption{For each CMH$_c$, $c\in \{0.1, 0.5, 1, 1.5, 2, 2.5, 3\}$, observed M-H acceptance rates are reported alongside estimates $\widehat{\text{ESJDR}}$ 
 relative to $\widehat{\text{ESJD}}_{GS} = 1.505$ and $\widehat{\text{MSER}}$ relative to $\widehat{\text{MSE}}_{GS} = 0.0214$.
 %, and  $\widehat{\text{ARE}}$ relative to $\hat{v}(X_1, P_{GS}) = 23.23$.  
 When possible, standard errors are given in parentheses.
}
\vspace{.1in}
 \label{tab1}
\begin{tabular}{|c||c|c|c|c|c|c|c|}
\hline\noalign{\smallskip}
c    & 0.1 & 0.5 & 1 & 1.5 & 2 & 2.5 & 3\\
\noalign{\smallskip}\hline  \hline \noalign{\smallskip}
$\widehat{\text{ESJDR}}$& 1.02  & 1.14 & 1.29  & 1.37 & 1.33  & 1.14 & 0.79  \\
 &  (0.003) & (0.003) & (0.003) &  (0.003) &  (0.004) &  (0.006) &  (0.011) \\
\noalign{\smallskip}\hline \noalign{\smallskip}
$\widehat{\text{MSER}}$         &  0.99    & 0.91  & 0.82  & 0.75  & 0.90  & 1.24   & 2.40   \\
&  (0.06)   &  (0.06) &  (0.05) &  (0.04) &  (0.06) &  (0.08)  &  (0.16)  \\
\noalign{\smallskip}\hline \hline \noalign{\smallskip}
%$\widehat{\text{ARE}}$  &         0.86   & 0.85 & 0.78 & 0.72 & 0.73 & 1.21 & 3.05 \\
%\noalign{\smallskip}\hline  \noalign{\smallskip}
Accept Rate & 0.99 &  0.91 & 0.75 & 0.58 & 0.41 & 0.27 & 0.18 \\
\noalign{\smallskip}\hline
\end{tabular}
\end{table}

Using similar simulation procedures, we compare GS and CMH$_q$ for fixed densities of sizes $q \in \{0.05, 0.10, 0.25, 0.50, 0.75, 0.90\}$.  The results exhibit familiar patterns  (Table \ref{tab2}).  Mainly, CMH$_q$ is more efficient or competitive with GS when $q$ is sufficiently small ($q \le 0.5$) but, as expected, suffers when the $q$-density neighborhoods are too large ($q > 0.5$).  Further, a comparison of Tables \ref{tab1} and \ref{tab2} suggests that even though CMH$_q$ enjoys larger average per iteration movement, the typical CMH$_c$ produces better estimates of $\beta$.  That is, a fixed width CMH neighborhood strategy is more efficient than a fixed density strategy with respect to estimation in this Normal-Normal setting.

\begin{table}[H] \centering
 \caption{For each CMH$_q$, $q\in \{0.05, 0.10, 0.25, 0.50, 0.75, 0.90\}$,  we present estimates $\widehat{\text{ESJDR}}$ 
 relative to $\widehat{\text{ESJD}}_{GS} = 1.505$ and $\widehat{\text{MSER}}$ relative to $\widehat{\text{MSE}}_{GS} = 0.0214$.
 %, and  $\widehat{\text{ARE}}$ relative to $\hat{v}(X_1, P_{GS}) = 23.23$.  
Standard errors are given in parentheses.
}
\vspace{.1in}
 \label{tab2}
\begin{tabular}{|c||c|c|c|c|c|c|}
\hline\noalign{\smallskip}
q    & 0.05 & 0.10 & 0.25 & 0.50 & 0.75 & 0.90 \\
\noalign{\smallskip}\hline  \hline \noalign{\smallskip}
$\widehat{\text{ESJDR}}$& 1.05  & 1.11 & 1.36  & 2.05 & 3.60  & 6.19  \\
 &  (0.003) & (0.003) & (0.003) &  (0.004) &  (0.007) &  (0.011)  \\
\noalign{\smallskip}\hline \noalign{\smallskip}
$\widehat{\text{MSER}}$         &  1.00    & 0.96  & 0.96  & 0.89  & 1.34  & 1.96     \\
&  (0.06)   &  (0.06) &  (0.06) &  (0.06) &  (0.08) &  (0.12)   \\
%\noalign{\smallskip}\hline \hline \noalign{\smallskip}
%$\widehat{\text{ARE}}$    & 0.88 & 0.87 & 0.89 & 0.87 & 1.21 & 1.87 \\
\noalign{\smallskip}\hline
\end{tabular}
\end{table}

\subsection{Bayesian random effects model}
Consider the following special case of the Bayesian version of the random effects model presented in \cite{john:jone:2013}. 
% studied by \citet{jone:hobe:2001} and \citet{hobe:geye:1998}.  
Let $Y_{ij}$ represent the $j$th observation on subject $i$ where $i \in \{1,\ldots,K\}$ and $j \in \{1,  \ldots, m\}$.  Then for $\theta=(\theta_1,\ldots,\theta_K)^T$ and $\lambda = (\lambda_e,\lt)^T$,
\begin{equation}\label{eq:mod1}
\begin{split}
Y_{i,j} | \theta, \mu, \lambda & \stackrel{ind}{\sim} N(\theta_i, \lambda_e^{-1}) \\
\theta_i | \mu, \lambda & \stackrel{iid}{\sim} N(\mu,\lt^{-1})\\
\mu & \sim N(m_0, s_0^{-1})) \\
\lambda_\theta & \sim \text{Gamma}(a_1, b_1) \\
\lambda_e & \sim \text{Gamma}(a_2, b_2) \\
\end{split}
\end{equation} 
where $m_0$ and $s_0$ are assumed known and we say $X \sim \text{Gamma}(a, b)$ if it has density proportional to $x^{a-1} e^{-b x}$.
Letting $y = \{y_{ij}\}$ represent the vector of observed data, the corresponding posterior distribution $\varpi$ is characterized by density 
$\pi(\theta,\mu,\lam | y) \propto \pi(y | \theta, \mu, \lam) \pi(\theta | \mu, \lam) \pi(\mu) \pi(\lam)$
with support $\mathcal{X} = \mathbb{R}^{K+1} \times \mathbb{R}_+^2$ and where the $\pi$ represent the densities defined by \eqref{eq:mod1}.  
Further, suppressing dependence on $y$, the full conditional densities $\pi(\theta|\mu,\lam)$, $\pi(\mu|\theta,\lam)$  and $\pi(\lam|\mu,\theta)$ are defined by the following full conditional distributions:
\[
\begin{split}
\theta_i | \mu, \lambda & \stackrel{ind}{\sim} N\left( \frac{\lt \mu + m\lame\overline{y}_i}{\lt + m\lame}, \; \frac{1}{\lt + m\lame}\right) \; \; \text{ for } i \in \{1,\ldots, K\} \\
\mu | \theta, \lambda  & \sim N\left( \frac{s_0m_0 + K\lt \overline{\theta}}{s_0 + K\lt}, \; \frac{1}{s_0 + K\lt}\right) \\
\lt | \theta,\mu & \sim \Gamma\left(\frac{K}{2} + a_1, \; \; \frac{\sum_{i=1}^K (\theta_i - \mu)^2}{2} + b_1 \right) : = \Gamma\lp\alpha_1,\beta_1(\theta,\mu)\rp  \\
\lambda_e | \theta, \mu & \sim \Gamma \left( \frac{Km}{2} + a_2, \;\; \frac{\sum_{i=1}^K m (\theta_i - \overline{y}_i)^2 + \text{SSE}}{2} + b_2\right)  : = \Gamma\lp\alpha_2,\beta_2(\theta,\mu)\rp  \\\\
\end{split}
\]
where $\overline{\theta} = K^{-1}\sum_{i=1}^K \theta_i$, $\overline{y}_i = m^{-1} \sum_{j=1}^{m} y_{ij}$ and $\text{SSE} = \sum_{i,j} (y_{ij} - \overline{y}_i)^2$.

\begin{comment}

K=3
m=10
a1=a2=30
d1 = 1/(2*a1+K-2)
d2 = 1/(2*a2+K*m-2)
p3 = min(1/(1+m*K*d1*d2), m/abs(m*(1+m*K*d1*d2) - 2*m*d1*(K+m)))
p3
p3=1/3
p2 = m/K*p3
p1 = m*(1-p3*(1+m*K*d1*d2))/(2*K*m*d1)
p1
p2
delta = matrix(0, 50,50)
A4 = seq((2+m*d2)/m, 2/(K*m*d1), len=50)
for(i in 1:50){
A1 = seq(K*m*A4[i], 2/d1, len=50)
for(j in 1:50){
delta[i,j] = max(2/3, 2/3 + 1/3*K*m*A4[i]/A1[j], 2/3 + 1/3*K/m, 1/3 + 1/3*d1*A1[j], 2/3 + 1/3*2/m/A4[i] + 1/3*d2/A4[i])
}
}
> min(delta)
[1] 0.7666667
> 23/30
[1] 0.7666667
> 

sum(rowSums(delta == min(delta))*c(1:50))
sum(colSums(delta == min(delta))*c(1:50))

A4try = A4[8]
A1try = seq(K*m*A4try, 2/d1, len=50)[28]
> A4try
[1] 0.7621212
> A1try
[1] 77.4898

A4try = 0.75
A1try = 75
> max(2/3, 2/3 + 1/3*K*m*A4try/A1try, 2/3 + 1/3*K/m, 1/3 + 1/3*d1*A1try, 2/3 + 1/3*2/m/A4try + 1/3*d2/A4try)
[1] 0.7666667

qmax < (1-min(delta))/2 = 7/60

\end{comment}

Given the intractable nature of $\pi(\theta, \mu,\lam | y)$, posterior inference requires MCMC methods.  To this end, we compare the GS and CMH using data $y$ simulated from \eqref{eq:mod1} with $K=3$, $m=10$, $m_0=0$, $s_0=1$, and $a_1=b_1=a_2=b_2=2$.   Assuming the true nature of this data is
unknown, we apply the GS and CMH under the hyperparameter setting  $a_1=b_1=a_2=b_2 = 30$. Further, for both algorithms, we use starting values of $\lp \theta^{(0)}, \mu^{(0)}, \lambda^{(0)} \rp = \lp (\overline{y}_1, \overline{y}_2, \overline{y}_3), 0, (1,1)\rp$ and update $\theta$, $\mu$, $\lambda$ with equal ($1/3$) probability in each iteration.  Throughout, we implement the CMH using neighborhoods for which size increases relative to the associated full conditional standard deviation.  Specifically, for $\vareps_\theta, \vareps_\mu, \vareps_\lambda > 0$,
$B(\theta|\mu,\lam)\subset \mathbb{R}^K$ is a sphere centered at $\theta$ with radius $\varepsilon_\theta\sqrt{\frac{1}{\lt + m\lame}}$,  
$B(\mu | \theta, \lam) = \mu \pm \varepsilon_\mu \sqrt{\frac{1}{s_0 + K\lt}} \subset \mathbb{R}$, and $B(\lam|\theta,\mu) = B_{\lt}(\lt|\theta,\mu) \times B_{\lame}(\lame|\theta,\mu) \subset \mathbb{R}^2_+$  is a rectangle centered at $\lam$ with  
\[
B_{\lt}(\lt|\theta,\mu) = \lt \pm \varepsilon_\lambda \frac{\sqrt{\alpha_1}}{\beta_1(\theta,\mu)}  \;\; \; \text{ and }\;\;\; B_{\lame}(\lame|\theta,\mu) = \lame \pm   \varepsilon_\lambda  \frac{\sqrt{\alpha_2}}{\beta_2(\theta,\mu)} \; .
\]
%Reflecting the dependence of $\text{Var}(\theta|\mu,\lambda)$ and $\text{Var}(\mu|\theta,\lambda)$ on $\lambda$ and the dependence of $\text{Var}(\lambda|\theta,\mu)$ on $(\theta,\mu)$, notice that the widths of the neighborhoods vary with the current values of $(\theta,\mu,\lambda)$. 
There are many reasonable strategies for selecting neighborhood parameters $(\vareps_\theta, \vareps_\mu, \vareps_\lambda)$.  Here, we choose sets that produce similar acceptance rates for the individual $(\theta,\mu,\lambda)$ updates. 

The GS and CMH are both Harris ergodic and Feller.  Further, under our chosen hyperparameter settings,  the results of \cite{john:jone:2013} guarantee geometric ergodicity for the GS with drift condition $P_{GS}V(\mu,\theta,\lambda) \le \gamma V(\mu,\theta,\lambda) + b$ for $b < \infty$, $\gamma = 23/30$, and 
 \[
V(\mu,\theta,\lam) =  75 \lt^{-1} + \lame^{-1} + \sum_{i=1}^K (\theta_i - \mu)^2+0.75 \sum_{i=1}^K m_i (\theta_i - \overline{y}_i)^2 + e^{\lt} + e^{\lame}+10 (\mu-\ooy)^2 +  \frac{K\lt}{s_0 + K\lt }(\overline{\theta}-\ooy)^2 
\]
where $\overline{\overline{y}} = K^{-1} \sum_{i=1}^K \overline{y}_i$. 
  In turn,  Proposition \ref{prop:re} follows from Theorem \ref{thm:gscmh}.
% \[
% \delta_1 = \frac{1}{61} \; ,
% \hspace{.2in}  \delta_2 = \frac{1}{88} \; , 
% \hspace{.2in} \text{ and } \hspace{.2in}  \delta_3 = \frac{189}{249} \; .
% \]
%Letting $p_1=p_2=p_3=1/3$, $K=3$, $m=10$ we can derive $A_1=75$, $A_4=0.75$
%In the notation of \cite{john:jone:2013}, let 
%\[
%\begin{split}
%\gamma = \frac{2}{3}  + \frac{1}{3}\max\{ & 0.3, \; 0.3, \; \frac{14}{61},  \; \frac{279}{990}, \;  0.3 \} \\
%\end{split}
%\]
\begin{proposition}\label{prop:re}
 %$q_{\max} < 7/60$
 The CMH is geometrically ergodic if $\varepsilon_\theta \le 0.6567$, $\varepsilon_\mu \le 0.1467$, and $\varepsilon_\lambda \le 0.0009$.
% under the following neighborhood restrictions:
% \begin{enumerate}
%  \item 
% $\varepsilon_\theta < 0.0487$, the 0.5194 quantile of the standard Normal distribution; 
% \item 
% $\varepsilon_\mu < 0.1467$, the 0.5583 quantile of the standard Normal distribution; and
% \item $\varepsilon_\lambda < 2.1586$, that is, 
% \[
% 2*\varepsilon_\lambda \frac{\sqrt{a_i}}{b_i} =  2*\varepsilon_\lambda \frac{\sqrt{30}}{30}
% \] 
% is less than the 0.1167 quantile of the Gamma$\lp 30, 30 \rp$ distribution.
% %$2^{-1}(1-\gamma) = 7/60$ 
% \end{enumerate}
% 
\end{proposition}

To compare the efficiency of the GS and CMH, we focus on their estimation of  posterior expectation $\beta=E(\mu|y)$.   
To begin, we ran $N=1000$ independent, length $n=1000$ runs of the GS and the CMH under each set of $(\vareps_\theta, \vareps_\mu, \vareps_\lambda)$ listed in Table \ref{tab3}, the first of which meets the conditions of Proposition \ref{prop:re}.
In each CMH setting, we obtained estimates $\widehat{\text{ESJDR}}$ and $\widehat{\text{MSER}}$ (Table \ref{tab3}).  Since the true value of $\beta$ is unknown, $\widehat{\text{MSER}}$ was calculated from
\[
\widehat{\text{MSE}}_{CMH} =
\frac{1}{N}\sum_{i=1}^{N}(\hat{\beta}_{CMH,n}^{(i)} - \beta^*)^2 
\hspace{.2in} \text{ and } \hspace{.2in}
\widehat{\text{MSE}}_{GS} =
\frac{1}{N}\sum_{i=1}^{N}(\hat{\beta}_{GS,n}^{(i)} - \beta^*)^2
\]
where $\beta^*$ is an independent estimate of $\beta$ based on a GS run of length $10^6$.
In addition, we estimated the 
overall CMH acceptance rates  from independent runs of length $n=10^5$ for each set $(\vareps_\theta, \vareps_\mu, \vareps_\lambda)$.  In each case, the component-wise acceptance rates of $(\theta,\mu,\lambda)$ are within 0.008 of the overall rate.

These simulation results bolster the observations made in the toy Normal-Normal setting.  Mainly, for reasonably sized neighborhoods, the CMH is more efficient than the GS in both its exploration of the state space and in its estimation of $\beta$.  We also observe a similar phenomena regarding the optimal neighborhood size;  the neighborhood setting that yields the most efficient CMH estimates (i.e.~smallest MSE) corresponds to the setting that facilitates the most efficient movement around the state space (i.e.~largest ESJD).  Perhaps coincidentally,  the optimal CMH settings for the Normal-Normal and random effects models also both have corresponding CMH acceptance rates of roughly 60 percent.

\begin{table}[H]\centering
 \caption{For the CMH under each given set of $(\varepsilon_\theta,\vareps_\mu,\vareps_\lambda)$ values, observed M-H acceptance rates are reported alongside estimates $\widehat{\text{ESJDR}}$ 
 relative to $\widehat{\text{ESJD}}_{GS} = 0.443$ and $\widehat{\text{MSER}}$ relative to $\widehat{\text{MSE}}_{GS} = 0.002$.
 %, and  $\widehat{\text{ARE}}$ relative to $\hat{v}(\mu, P_{GS}) = 1.94$.  
 When possible, standard errors are given in parentheses.
}
\vspace{.1in}
 \label{tab3}
\begin{tabular}{|c||c|c|c|c|c|c|}
\hline\noalign{\smallskip}
$\vareps_\theta$ & 0.65 & 1.3 & 1.7 & 2.3 & 3.0 & 3.9 \\
$\vareps_\mu$ & 0.14 & 0.5 & 0.9 & 1.4 & 2.0 & 2.9 \\
$\vareps_\lambda$ & 0.0009 &  0.9 & 1.3 & 1.9 &  2.6 & 3.4 \\
\noalign{\smallskip}\hline  \hline \noalign{\smallskip}
$\widehat{\text{ESJDR}}$& 1.01 & 1.06  & 1.11 & 1.15  & 1.08 & 0.85\\
&(0.003)  &  (0.003) & (0.003) & (0.003) &  (0.004) & (0.008) \\
\noalign{\smallskip}\hline \noalign{\smallskip}
$\widehat{\text{MSER}}$         & 0.86  & 0.82    & 0.76  & 0.69  & 0.92 & 7.74\\
& (0.05) &  (0.05)   &  (0.05) &  (0.04) &  (0.05)  & (0.45) \\
\noalign{\smallskip}\hline \hline \noalign{\smallskip}
%$\widehat{\text{ARE}}$  &        0.92 &  0.73   & 0.78 & 0.91 & 5.14 \\
%\noalign{\smallskip}\hline  \noalign{\smallskip}
Accept Rate & 0.99 & 0.90 & 0.80 &  0.60 & 0.40 & 0.20 \\
\noalign{\smallskip}\hline
\end{tabular}
\end{table}

\section{Appendix}

\subsection{Proof of Theorem \ref{thm:gscmh}}

By assumption, the GS is geometrically ergodic with drift function $V: \mathcal{X} \to [1,\infty)$ that is unbounded off compact sets and satisfies the following drift condition for $0 < \gamma < 1$ and $b < \infty$:
\[
P_{GS}V(x) :=  \sum_{i=1}^m p_i P_{GS_i}V(x) \le \gamma V(x) + b 
\]
where $P_{GS_i}V(x) =  \int V\left(x_{[i]}, x_i', x^{[i+1]}\right) \pi(x_i'|x_{-i}) \mu_i(dx_i')$.
To extend these results to the CMH, 
we will establish the following drift condition.  Geometric ergodicity follows directly.
Define function $\tilde{V}:\mathcal{X} \to [1,\infty)$
\[
\tilde{V}(x) = V(x) + aW(x) 
\]
for
\[
\begin{split}
W(x) & = \max_j\lb \left(V(x) - V(x_{[j]},x_j',x^{[j+1]})\right) I(x_j' \in B_j(x_j|x_{-j}))\rb  \\
a & \in \left( \frac{q_{\max}}{1-2q_{\max}} , \;  \frac{(1-q_{\max}) - \lambda(1-q_{\min})}{\lambda(1-q_{\min}) - q_{\min}(1-q_{\max})}\right) \\
\end{split}
\]
where $0 \le W(x) \le V(x)$ and the interval for $a$ is guaranteed to be non-empty under the restrictions on $q_{\min},q_{\max}$ and assuming, without loss of generality, that $\gamma > q_{\min}(1-q_{\max})(1-q_{\min})^{-1}$.
Then $\tilde{V}$ is unbounded off compact sets on $\mathcal{X}$ and satisfies the following drift condition:
\begin{equation}\label{eq:proofdrift}
\begin{split}
P_{CMH}\tilde{V}(x) & \le \tilde{\gamma} \tilde{V}(x) + (a+1)b\\
\end{split}
\end{equation}
where
\[
\tilde{\gamma} = \max\lb (a+1)\left( \frac{\gamma }{1-q_{\max}}  - \frac{q_{\min}}{1-q_{\min}} \right), \; 
\frac{a+1}{a}\frac{q_{\max}}{1-q_{\max}}\rb\;
\]
and $0 < \tilde{\gamma} < 1$ by the definition of $a$.   To establish the drift condition at \eqref{eq:proofdrift}, first define
 \[
\begin{split}
m_i(x,x_i') 
& = \frac{1}{\max\left\lbrace\; \int_{B_i^c(x_i|x_{-i})}\pi(z_i|x_{-i}) \mu_i(dz_i), \; \int_{B_i^c(x_i'|x_{-i})}\pi(z_i|x_{-i}) \mu_i(dz_i) \right\rbrace} \;\; \text{ for } x,x' \in \mathcal{X} \\
 \end{split}
 \]
and notice that
\[
\frac{1}{1-q_{\min}} \le m_i(x,x_i') \le \frac{1}{1-q_{\max}} 
\]
and $q_i(x_i'|x) \alpha_i(x_i'|x) = \pi(x_i' | x_{-i}) m_i(x,x_i') I(x_i' \in B_i^c(x_i|x_{-i}))$.
It follows that
\[
\begin{split}
P_{CMH_i}V(x) 
& = \int V(x_{[i]},x_i',x^{[i+1]})   q_i(x_i'|x) \alpha_i(x_i'|x)  \mu_i(dx_i')   + V(x) \left[1 - \int q_i(x_i'|x) \alpha_i(x_i'|x)  \mu_i(dx_i')  \right]  \\
& = \int \limits_{B_i^c(x_i|x_{-i})}V(x_{[i]},x_i',x^{[i+1]})   \pi(x_i' | x_{-i}) m_i(x,x_i') \mu_i(dx_i')   + V(x) \left[1 - \hspace{-.2in}\int\limits_{ B_i^c(x_i|x_{-i})} \hspace{-.2in}\pi(x_i' | x_{-i}) m_i(x,x_i') \mu_i(dx_i')  \right]  \\
&  = \int V(x_{[i]},x_i',x^{[i+1]}) \pi(x_i' | x_{-i})m_i(x,x_i') \mu_i(dx_i')   - V(x)\left[ \int \pi(x_i' | x_{-i}) m_i(x,x_i')\mu_i(dx_i')- 1\right] \\
& \hspace{.2in} + \int\limits_{B_i(x_i|x_{-i})} (V(x) -  V(x_{[i]},x_i',x^{[i+1]}))\pi(x_i' | x_{-i}) m_i(x,x_i') \mu_i(dx_i')  \\ 
&  \le \frac{1}{1-q_{\max}} \int V(x_{[i]},x_i',x^{[i+1]}) \pi(x_i' | x_{-i}) \mu_i(dx_i') - V(x)\left[ \frac{1}{1-q_{\min}}- 1\right] \\
& \hspace{.2in} + W(x) \int\limits_{B_i(x_i|x_{-i})}  \pi(x_i' | x_{-i})m_i(x,x_i') \mu_i(dx_i') \\
&  \le \frac{1}{1-q_{\max}}P_{GS_i}V(x)  - \frac{q_{\min}}{1-q_{\min}} V(x)+ \frac{q_{\max}}{1-q_{\max}}W(x) \\
\end{split}
\]
so that
\[
\begin{split}
P_{CMH}V(x) & = \sum_{i=1}^m p_i P_{CMH_i}V(x) \\
& \le \frac{1}{1-q_{\max}} \sum_{i=1}^m p_i P_{GS_i}V(x) - \frac{q_{\min}}{1-q_{\min}} V(x) + \frac{q_{\max}}{1-q_{\max}}W(x) \\
& = \frac{1}{1-q_{\max}} P_{GS}V(x)    - \frac{q_{\min}}{1-q_{\min}} V(x) + \frac{q_{\max}}{1-q_{\max}}W(x) \\
& \le \left( \frac{\gamma }{1-q_{\max}}  - \frac{q_{\min}}{1-q_{\min}} \right) V(x)  +\frac{q_{\max}}{1-q_{\max}}W(x) + b  \; .\\
\end{split}
\]

%Consider the first case in which there exists $c \in \mathbb{R}^+$ such that
%\[
%W(x) \le \sup_{x,x' \in \mathcal{X}, i\in \{1,\ldots,m\}} \; \lb \; \vline \; V(x) - V\left(x_{[i]},x_i',x^{[i+1]}\right) \; \vline I(x_i' \in B_i(x_i|x_{-i}))\rb \le c \; .
%\]
%Then geometric ergodicity of the CMH is guaranteed by the drift condition 
%\[
%\begin{split}
%P_{CMH}V(x) 
%& \le \left( \frac{\lambda }{1-q_{\max}}  - \frac{q_{\min}}{1-q_{\min}} \right) V(x)  + constant \;\\
%\end{split}
%\]
%where the assumptions on $q_{\min}, q_{\max}$ guarantee
%\[\frac{\lambda }{1-q_{\max}}  - \frac{q_{\min}}{1-q_{\min}} < 1 \; .\]

%On the other hand, consider the second case in which $(1-q_{\max})/(1-q_{\min}) > \lambda$, $q_{\max} < \frac{1}{2}$, and $q_{\min}q_{\max} + \lambda q_{\min} - 2q_{\max} + (1-\lambda) > 0$. 
Thus \eqref{eq:proofdrift} follows:
\[
\begin{split}
P_{CMH}\tilde{V}(x) 
& = P_{CMH} V(x) + aP_{CMH}W(x) \\
& \le (a+1)P_{CMH} V(x) \\
& \le  (a+1)\left( \frac{\gamma }{1-q_{\max}}  - \frac{q_{\min}}{1-q_{\min}} \right) V(x)  + \frac{a+1}{a}\frac{q_{\max}}{1-q_{\max}}aW(x) + (a+1)b\\
& \le \tilde{\gamma} \tilde{V}(x) + (a+1)b \; .\\
\end{split}
\]

\subsection{Proof of Theorem \ref{thm:subgeo}}

Since the GS and CMH are reversible with respect to $\pi$, we are able to prove this result using a {\it capacitance} argument. 
In general, let $\Phi$ be a reversible Markov chain with kernel $P$ and  let $P_0$ denote the restriction of $P$ to $L_{0,1}^2(\pi) = \{f \in L^2(\pi): \; E_\varpi f =0 \text{ and } E_\varpi f^2=1\}$.  The spectral radius of $P_0$ is  $r(P_0) = \sup\{|\lambda|: \lambda \in \sigma(P_0)\}$
where $\sigma(P_0) \subset [-1,1)$ is the spectrum of $P_0$.  Further, \cite{sinc:1992} establish that
\[
1 - 2\kappa \le r(P_0) \le 1 - \frac{\kappa}{2} 
\]
for capacitance $\kappa$:
\[
\kappa : = \inf_{S: 0 < \pi(S) \le 1/2} \frac{1}{\pi(S)} \int_S P(x,S^c) \pi(x) \mu(dx)
\]
where $\pi(S) = \int S(x)\pi(x) \mu(dx)$.  It is known that $\Phi$ is geometrically ergodic if and only if $r(P_0) < 1$ or, equivalently, $\kappa>0$.

Consider the CMH with $B_i(x_i|x_{-i})$ that satisfy \eqref{eq:nbhd}.  Thus there exists some  $0 < q_{\max} < 1$ for which
\[
\int_{B_i(x_i|x_{-i})} \pi(z_i|x_{-i}) \mu_i(dz_i) \le q_{\max} \; \text{ for all } x \in \mathcal{X}, \; i \in \{1,\ldots,m\} \; .
\]
By assumption, the CMH is geometrically ergodic so that 
\[
\kappa_{CMH} := \inf_{S: 0 < \pi(S) \le 1/2} \frac{1}{\pi(S)} \int_S P_{CMH}(x,S^c) \pi(x) \mu(dx) > 0\; .
\]
Geometric ergodicity of the GS will follow from establishing that $\kappa_{GS} \ge \kappa_{CMH} > 0$ where 
\[
\kappa_{GS} := \inf_{S: 0 < \pi(S) \le 1/2} \frac{1}{\pi(S)} \int_S P_{GS}(x,S^c) \pi(x) \mu(dx) \; .
\]
To this end, note that the CMH can only move from state $x \in S$ to $x' \in S^c$ when the M-H proposal is accepted.  That is, for $x \in S$
\[
\begin{split}
P_{CMH_i}(x, S^c)
& =  \int\limits_{\{x_i': (x_{[i-1]},x_i',x^{[i+1]}) \in S^c\}} q_i(x_i'|x) \alpha_i(x_i'|x) \mu_i(dx_i') \\
& \le \int\limits_{\{x_i': (x_{[i-1]},x_i',x^{[i+1]}) \in S^c\}}q_i(x_i'|x) \mu_i(dx_i') \\
& \le \frac{1}{1-q_{\max}} \int\limits_{\{x_i': (x_{[i-1]},x_i',x^{[i+1]}) \in S^c\}} \pi(x_i'|x_{-i})\mu_i(dx_i') \\
& = \frac{1}{1-q_{\max}} P_{GS_i}(x_{-i},S^c)\; .\\
\end{split}
\]
Finally, it follows that $P_{CMH}(x,S^c) \le \frac{1}{1-q_{\max}} P_{GS}(x,S^c)$ for $x \in S$ and, in turn,
\[
\begin{split}
0 < \kappa_{CMH} &  \le \frac{1}{1-q_{\max}} \kappa_{GS}\; .\\
\end{split}
\]

\subsection{Proof of Proposition \ref{prop:nn}}

First, consider CMH$_q$ with neighborhoods $B_i(x_i|x_{-i})$ having fixed mass $q$.  By setting $q_{\min}=q_{\max}=q$ in \eqref{eq:q}, geometric ergodicity for the CMH$_q$ is guaranteed when $q < \min\lb\frac{1}{2}, \; 1- \gamma \rb = 0.25$.
Next, consider CMH$_c$.  Recall that $X_i|X_{-i} \sim N(X_{-i}, \sig_i^2)$ where $\sig_1^2=1$ and $\sig_2^2=1/2$.  Thus the neighborhoods $B_i(x_i|x_{-i}) = (x_i - c\sig_i, x_i + c\sig_i)$ have mass
\[\begin{split}
\int_{B_i(x_i|x_{-i})} \pi(z_i|x_{-i}) dz_i & = \int_{x_i-c\sig_i}^{x_i+c\sig_i} \pi(z_i|x_{-i}) dz_i = \int_{\frac{x_i-x_{-i}}{\sig_i}-c}^{\frac{x_i-x_{-i}}{\sig_i}+c} \phi(z_i) dz_i  \\
\end{split}
\]
where $\phi(\cdot)$ represents the standard Normal density function. 
 In this case,
\[\begin{split}
q_{\min} := 0 & \le \int_{B_i(x_i|x_{-i})} \pi(z_i|x_{-i}) dz_i \le  \int_{ - c}^{c} \phi(z_1)dz_1 := q_{\max} \; \\
\end{split}
\]
where, by \eqref{eq:q}, geometric ergodicity for the CMH is guaranteed when
\[
q_{\max} < \frac{1-\gamma}{2} = 0.125 \; .
\]
Since $q_{\max} < 0.125$ for all CMH$_c$ with $c < 0.1573$, the result holds.

\subsection{Proof of Proposition \ref{prop:re}}

Recall that the GS drift condition holds with drift rate $\gamma = 23/30$.  Thus \eqref{eq:q} requires that CMH neighborhoods $B(\theta|\mu,\lambda)$, $B(\mu|\theta,\lambda)$, and $B(\lambda|\theta,\mu)$ each have a maximum density of 
\[
q_{\max} < \frac{1-\gamma}{2} = \frac{7}{60} \; 
\]
for all $(\theta,\mu,\lambda)$.  To this end,  first consider $B(\mu | \theta, \lam)= \mu \pm \varepsilon_\mu \sqrt{\frac{1}{s_0 + K\lt}}$.
Since $\mu|\theta,\lambda$ is Normal, an argument similar to that in the proof of Proposition \ref{prop:nn} shows that 
\[\begin{split}
&\int_{B(\mu|\theta,\lambda)} \pi(\mu'|\theta,\lambda) d\mu' \le  \int_{ - \vareps_\mu}^{\vareps_\mu} \phi(z)dz < q_{\max} \; \\
\end{split}
\]
so long as $\vareps_\mu < 0.1467$.  Next, consider $B(\theta | \mu, \lam) \subset \mathbb{R}^K$ and notice that 
$B(\theta | \mu, \lam) \subset B_1(\theta_1 | \mu, \lam) \times \cdots \times B_K(\theta_K | \mu, \lam)$ where $B_i(\theta_i | \mu, \lam) = \theta_i \pm \vareps_\theta  \sqrt{\frac{1}{\lt + m\lame}} \subset \mathbb{R}$.  Further, since
the $\theta_i | \mu,\lambda$ are independently Normal with variance $(\lt + m\lame)^{-1}$, we can write the joint  conditional density as $\pi(\theta|\mu,\lambda) = \prod_{i=1}^K \pi_i(\theta_i|\mu,\lambda)$. 
Thus using similar arguments to those above, $\vareps_\theta < 0.6567$ guarantees 
\[\begin{split}
\int_{B(\theta|\mu,\lambda)} \pi(\theta'|\mu,\lambda) d\theta'
& \le \prod_{i=1}^K \int_{B_i(\theta_i|\mu,\lambda)} \pi_i(\theta_i'|\mu,\lambda) d\theta_i'  \le\lp \int_{ - \vareps_\theta}^{\vareps_\theta} \phi(z)dz\rp^K < q_{\max} \; .\; \\
\end{split}
\]
Finally, consider
$B(\lambda | \theta, \mu) = B_{\lt}(\lt | \theta, \mu) \times B_{\lame}(\lame | \theta,\mu)$.
%and notice that $B_{\lt}(\lt|\theta,\mu) \subset B_{\lt}(\lt)$ and $B_{\lame}(\lame|\theta,\mu) \subset B_{\lame}(\lame)$ where
%\[
%B_{\lt}(\lt) =  \lt \pm \vareps_\lambda \frac{\sqrt{\frac{K}{2} + a_1}}{b_1}  \;\; \; \text{ and }\;\;\; B_{\lame}(\lame) = \lame \pm   \varepsilon_\lambda \frac{\sqrt{\frac{Km}{2} + a_2}}{b_2} \; .
%\]
For ease of exposition, let $\beta_i := \beta_i(\theta,\mu)$ for $i=1,2$
and recall that, independently,
$\lt | \theta,\mu \sim \Gamma\lp\alpha_1,\beta_1\rp$ and $\lambda_e | \theta, \mu  \sim  \Gamma\lp\alpha_2,\beta_2\rp$ with joint density $\pi(\lambda|\theta,\mu) = \pi(\lt|\theta,\mu)\pi(\lame|\theta,\mu)$.  Further, define $\tilde{\lambda}_{\theta} = \beta_1 \lt \sim \text{Exp}(\alpha_1)$ with density $f(\tilde{\lambda}_\theta)$ and $\tilde{\lambda}_e = \beta_2 \lame \sim \text{Exp}(\alpha_2)$ with density $f(\tilde{\lambda}_e)$ where we say $X \sim \text{Exp}(a)$ if it has density proportional to $ e^{-a x}$ for $x > 0$.
Thus, for $\vareps_\lambda < 0.0009$, 
\[\begin{split}
\int_{B(\lambda|\theta, \mu)} \pi(\lambda'|\theta,\mu) d\lambda'
& = \int_{B_{\lt}(\lt|\theta, \mu)} \pi(\lt'|\theta,\mu) d\lt' \cdot \int_{B_{\lame}(\lame|\theta, \mu)} \pi(\lame'|\theta,\mu) d\lame'\\
& = \int_{\beta_1\lt - \vareps_\lambda \sqrt{\alpha_1}}^{\beta_1\lt + \vareps_\lambda \sqrt{\alpha_1}} f(\tilde{\lambda}_\theta') d\tilde{\lambda}_\theta' \cdot \int_{\beta_2\lame - \vareps_\lambda \sqrt{\alpha_2}}^{\beta_2\lame + \vareps_\lambda \sqrt{\alpha_2}} f(\tilde{\lambda}_e') d\tilde{\lambda}_e'\\
& \le \int_0^{2\vareps_\lambda \sqrt{\alpha_1}} f(\tilde{\lambda}_\theta') d\tilde{\lambda}_\theta' \cdot \int_0^{2\vareps_\lambda \sqrt{\alpha_2}}f(\tilde{\lambda}_e') d\tilde{\lambda}_e'\\
& < q_{\max}\\
\end{split}
\]
where the first inequality is guaranteed by the structure of the Exponential densities.
\begin{comment}
fun = function(eps){
pexp(2*eps*sqrt(31.5), 31.5) * pexp(2*eps*sqrt(45), 45) - 7/60
}
> uniroot(fun,interval=c(0.00000000001, 1))
$root
[1] 0.00090741
\end{comment}

%\vspace{.2in}
%{\bf PAPERS TO RE-READ}
%
%\begin{itemize}
%\item Liu of course
%\item Geyer and Mira
% \item Liu, Wong, \& Kong (Covariance structure of the Gibbs sampler with applications to comparisons of estimators and augmentation schemes) look at improvements in efficiency when blocking CGS.  Under non-reversibility, they establish this result by directly comparing operator norms, not through Peskun ordering.   
% \item Roy (Spectral Analytic Comparisons for Data Augmentation)
%\end{itemize}
\bibliographystyle{ims}
\bibliography{mcref}

\end{document}